\documentclass[prd,a4paper,showpacs,preprintnumbers,nofootinbib]{revtex4} 

\usepackage{epsfig}

\newcommand{\open}{{<\kern -0.3 em{\scriptscriptstyle )}}}

\newcommand{\nslash}{\kern 0.2 em n\kern -0.45em /}
\newcommand{\Pslash}{\kern 0.2 em P\kern -0.56em \raisebox{0.3ex}{/}}
\newcommand{\pslash}{\kern 0.2 em p\kern -0.4em /}
\newcommand{\kslash}{\kern 0.2 em k\kern -0.45em /}
\newcommand{\Sslash}{\kern 0.2 em S\kern -0.56em \raisebox{0.3ex}{/}}

\newcommand{\be}{\begin{eqnarray}}
\newcommand{\ee}{\end{eqnarray}}

\newcommand{\sumint}{\kern 0.2 em {\textstyle\sum} \kern -1.1 em \int}

\newcommand{\gsim}{\mbox{\raisebox{-0.6ex}{$\stackrel{>}{\sim}$}}\:}
\newcommand{\lsim}{\mbox{\raisebox{-0.6ex}{$\stackrel{<}{\sim}$}}\:}
\renewcommand{\d}{{\rm d}}

\begin{document} 

\title{Single transverse-spin asymmetries in forward pion production
at high energy: incorporating small-$x$ effects in the target}

\author{Dani\"el Boer}
\email{D.Boer@few.vu.nl}
\affiliation{Dept.\ of Physics and Astronomy, 
Vrije Universiteit Amsterdam, \\
De Boelelaan 1081, 1081 HV Amsterdam, The Netherlands}

\author{Adrian Dumitru}
\email{dumitru@th.physik.uni-frankfurt.de}
\affiliation{Institut f\"ur Theoretische Physik, 
J.~W.~Goethe Universit\"at\\
Max-von-Laue Strasse 1, D-60438 Frankfurt am Main, Germany}

\author{Arata Hayashigaki}
\email{aratah@th.physik.uni-frankfurt.de}
\affiliation{Institut f\"ur Theoretische Physik, 
J.~W.~Goethe Universit\"at\\
Max-von-Laue Strasse 1, D-60438 Frankfurt am Main, Germany}

\date{\today}

\begin{abstract}
We consider single-inclusive forward pion production in high-energy
proton-proton collisions at RHIC energies. A good baseline description
of the transverse momentum distributions at high rapidity is obtained
within Mueller's dipole formalism with an anomalous dimension
incorporating an ``extended geometric scaling'' window between the
saturation and DGLAP regimes. We then formulate pion production 
for transversely polarized projectiles within the same approach. We
assume that an azimuthal, spin-dependent asymmetry arises from the 
so-called Sivers effect and investigate the single transverse-spin 
asymmetry $A_N$ at 200~GeV and 500~GeV center-of-mass energy. A simple 
parameterization of the Sivers functions from the literature compares 
reasonably well with the high-energy STAR data if the overall normalization 
is scaled up by at least a factor of two. The STAR data might therefore 
indicate that the Sivers effect is significantly stronger than thought so far.
We also analyze higher-twist contributions to $A_N$ and find that they
largely cancel.
\end{abstract}


\maketitle


\section{Introduction}

Large single transverse-spin asymmetries (SSAs) in $p^\uparrow \, p
\to \pi \, X$ have been observed in fixed target experiments
\cite{ANdata} over the last 15 years. A non-vanishing SSA
implies that the azimuthal distribution of pions in the final state
depends on the direction of the transverse spin of the polarized
projectile proton. These observations pose a challenge for theorists
who want to explain such asymmetries in terms of quark and gluon
degrees of freedom.  It is clear that the asymmetries result from
spin-orbit couplings and chiral dynamics in low-energy QCD, but to
describe such couplings and chiral symmetry breaking within quantum
field theory is far from straightforward. Attempts to describe the
observable
\be A_N = \frac{\d\sigma(p^\uparrow \, p \to \pi \, X) -
\d\sigma(p^\downarrow \, p \to \pi \, X)}{\d\sigma(p^\uparrow \, p \to \pi
\, X) + \d\sigma(p^\downarrow \, p \to \pi \, X)}
\label{eqn:asymmetry}
\ee
perturbatively at leading twist and within the framework of collinear
factorization where partonic transverse momenta in hadrons are assumed
to play a subdominant role (because the associated scale is of order
$\Lambda_{QCD}$), result in very small asymmetries \cite{Kane}.
Several proposals that go beyond leading twist or collinear
factorization have been put forward
\cite{Sivers,Collins-93b,Anselmino,moremechanisms} and can in
principle account for the large asymmetries. They all
involve new polarization-dependent parton distribution or
fragmentation functions, which however cannot be calculated within
perturbative QCD (pQCD) because of the soft dynamics. So far, these
functions could also not be calculated from Euclidean QCD on a
lattice.  Hence, reliable predictions can only be made once the
unknown functions have been extracted from experiments. As a further
complication the evolution properties of the functions, needed to
describe the energy dependence of the asymmetry observables, are also
not known yet, because of the much more complicated operator structure
as compared to the ordinary parton densities.  Despite these
limitations, a fairly successful phenomenology has emerged.  Given the
large magnitude of the effects it is certainly worth exploring such
SSAs further, both experimentally and theoretically.

Due to the rather low energies and transverse momenta probed by fixed
target experiments, a factorized pQCD description of the process may
be somewhat questionable. Therefore, the high-energy collider
experiments with polarized protons at BNL's Relativistic Heavy-Ion
Collider (RHIC) were eagerly awaited. Recently, the STAR collaboration
at RHIC has found that the large single spin asymmetries persist at
collider energies of $\sqrt{s}= 200$~GeV and transverse momenta up to
$p_T\sim2.5$~GeV/c~\cite{STAR03}. This has been confirmed by the
BRAHMS experiment at RHIC~\cite{Brahms}. Ref.~\cite{Soffer} suggested
that the asymmetries at low and high center of mass energies are
actually two different phenomena, which was based on the observation
that only at RHIC energies the unpolarized inclusive pion
production cross section is described satisfactorily by a leading-twist,
next-to-leading order (NLO) pQCD calculation. The fact that this does
not work for fixed-target energies casts some doubt on the validity of
the factorization assumption for those cases.

In this paper we shall specifically address the high-energy collider
data on SSAs. As a first step, we attempt to obtain a good baseline
for inclusive pion production in unpolarized $pp$ collisions. SSAs
are usually measured at forward rapidities $y$ (in the hemisphere of
the polarized projectile) where they are found to be largest.  The
large rapidity coverage of the forward $\pi^0$ detectors at STAR
enable measurements of pion production up to $y\sim4$~\cite{Bland}. At
large $y$, however, the kinematics is very asymmetric: the hadrons in
the final state emerge from collisions of projectile partons with
large light-cone momentum fraction $x_1\sim 0.1$ - 1 with target
partons carrying a very small momentum fraction $x_2\ll1$. When
$\alpha_s\log\,1/x_2$ is of order 1, small-$x$ effects in the target
become important and modify its gluon distribution function as
compared to that obtained in the DGLAP approximation (specifically,
its anomalous dimension). The purpose of this paper is to show how
such small-$x$ effects can be accounted for in the description of SSAs
at forward rapidities and high energies.  Their relevance for deep
inelastic scattering (DIS) at HERA and forward pion production in
$p\,A$ collisions at RHIC has already been discussed in (among others)
refs.~\cite{GBW,Stasto:2000er,IIM} and~\cite{KKT,JK,Adrian1,Adrian2},
respectively.

We will assume that the asymmetry arises predominantly from the
so-called Sivers effect \cite{Sivers}. Within the framework of spin
and transverse momentum dependent distribution and fragmentation
functions, it has been shown \cite{Anselmino-04} that other possible
mechanisms, such as the Collins effect \cite{Collins-93b}, cannot
describe the asymmetries without other additional contributions, most
importantly the Sivers effect. We therefore simply restrict to the
Sivers effect, but inclusion of the other types of spin effects will
proceed along similar lines.

The first study of small-$x$ effects, more specifically of saturation
effects within the so-called ``Color Glass Condensate'' (CGC) formalism, 
on a single spin asymmetry concerned production of polarized 
$\Lambda$ baryons in collisions of unpolarized protons on 
nuclei~\cite{Boer:2002ij}.
No effects from small-$x$ quantum evolution, such as geometric
scaling violations (see below), were taken into account though. Also,
we shall show here that higher-twist saturation effects play no significant
role for $pp$ collisions at RHIC energy, as may be expected from estimates of
the saturation scale $Q_s$.

This paper is organized as follows.  In the next section we describe a
model for the Sivers effect using a well-known parameterization taken
from the literature.  In sec.~III-A we formulate single-inclusive {\em
forward} hadron production in unpolarized $pp$ collisions within the CGC 
formalism and show that
good agreement with the experimental data emerges. Next, by applying
the formalism to $p^\uparrow p$ collisions, in sec.~III-B we compare
our results for the SSA to recent STAR data at large rapidity ($\simeq
3.8$).  Sec.~IV is devoted to a discussion of the importance of
higher-twist effects and finally a summary is given in sec.~V.

\section{A model for the Sivers effect}

The Sivers effect is a correlation between the direction of the
transverse spin of the proton and the transverse momenta of its
unpolarized partons. The Sivers effect in the process $p^\uparrow \, p
\to \pi \, X$ has first been analyzed by Anselmino {\it et
al.}~\cite{Anselmino}. The Sivers functions for valence $u$ and $d$
quarks were extracted from a fit to fixed-target data under the
assumption that the asymmetry arises solely from this effect. That fit
relied on a calculation of both the polarized and unpolarized cross
sections within the LO DGLAP pQCD approach and partial inclusion of
partonic transverse momentum dependence. In the
present paper we will analyze whether the Sivers function
from~\cite{Anselmino} is consistent with the high-energy STAR data, when
incorporated in the CGC formalism. To
be precise, we shall use the Sivers functions presented in
refs.~\cite{AM98,ABM-99}. Although updated fits \cite{ADM} incorporate
more detailed transverse momentum dependence, the simpler
parameterizations of~\cite{AM98,ABM-99} will suffice for our present
purposes. The number densities of $q=u,\bar{u},d,\bar{d},\ldots$
partons in a polarized proton are defined as~\cite{ADM}
\be
f_{q/p^\uparrow} (x,\vec{k}_t) = f_{q/p}(x,k_t) + 
\frac{1}{2} \Delta^N f_{q/p^\uparrow}(x,k_t) 
\frac{\vec{S}_t \cdot(\vec{P} \times \vec{k}_t)}
{|\vec{S}_t| \, |\vec{P}| \, |\vec{k}_t|}~,
\label{eqn:number_density}
\ee
where $\vec{S}_t$ and $\vec{P}$ are the transverse polarization and
three-momentum vectors of the proton, respectively, and $\vec{k}_t$ is
the quark's intrinsic transverse momentum. The function
$f_{q/p}(x,k_t)$ is the density of quarks with transverse momentum
$k_t$ in an unpolarized proton, from which the familiar quark
distribution $f_{q/p}(x)$ arises by integration over $k_t$, see below.
$\Delta^N f_{q/p^\uparrow}(x,k_t)$ denotes the Sivers function (see
ref.\ \cite{Trento} for a comparison to another frequently used
notation $f_{1T}^\perp$ \cite{BM-98}).

\begin{figure}[hbt]
\centering
\centerline{\epsfig{figure=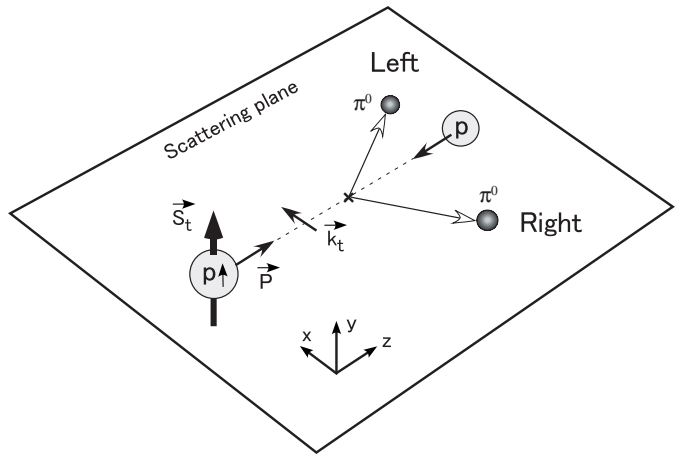,width=3.5in,angle=0}}
\caption{Illustration of the $p^\uparrow \, p \to \pi^0 \, X$ process.
The kinematics is simplified by selecting the pion production
plane ($x-z$) perpendicular to the polarization $\vec{S}_t=(0,S_y,0)$ of the
proton. Parton transverse momenta are taken to be (anti-) parallel to
the $x$-direction. We define left (right)-hand production of
pions in the direction of $\vec{S}_t\times \vec{P}>0\,(<0)$ to match
the experimental situation.}
\label{fig:kinematics}
\end{figure}
Fig.~\ref{fig:kinematics} illustrates the $p^\uparrow \, p \to \pi^0
\, X$ collision process, simplifying the kinematics by taking the
$\pi^0$ production plane ($x-z$) perpendicular to the polarization of
the proton, $\vec{S}_t=(0,S_y,0)$, where the Sivers effect is maximal.
The collision axis is along the $z$-direction, i.e.\
$\vec{P}=(0,0,P_z)$. Following the standard experimental definition,
we define the left and right directions of $\pi^0$ production in the
forward (positive hadron rapidity $y_h$) region from the perspective
of the polarized proton as indicated in the figure. This definition
implies that if $\Delta^N f_{q/p^\uparrow}(x,k_t) > 0$, then the quark
in the polarized proton has a preference for moving to the left, i.e.\
with $\vec{k}_t$ pointing in the positive $x$-direction.

The difference between upward and downward 
spin polarizations of the proton, which enters in the numerator of $A_N$, 
therefore picks up a leading contribution from the Sivers functions. 
Flipping the spin of the proton is equivalent to changing the sign of 
$\vec{k}_t$:
\be
\Delta^N f_{q/p^\uparrow}(x,\vec{k}_t) &\equiv&
f_{q/p^\uparrow}(x,\vec{k}_t)-f_{q/p^\downarrow}(x,\vec{k}_t)
= f_{q/p^\uparrow}(x,\vec{k}_t)-f_{q/p^\uparrow}(x,-\vec{k}_t)
\nonumber\\
&=& \Delta^N f_{q/p^\uparrow}(x,k_t) 
\frac{\vec{S}_t \cdot(\vec{P} \times \vec{k}_t)}
{|\vec{S}_t| \, |\vec{P}| \, |\vec{k}_t|}~,
\label{eqn:difference}
\ee 
where $\Delta^N f_{q/p^\uparrow}(x,\vec{k}_t)$ is an odd function
of $\vec{k}_t$. The Sivers effects cancel in the sum over polarization
states, which enters in the denominator of $A_N$ and yields (twice)
the leading-twist unpolarized function,
\be 
f_{q/p^\uparrow}(x,\vec{k}_t)+f_{q/p^\downarrow}(x,\vec{k}_t)=
2f_{q/p}(x,k_t).
\label{eqn:sum}
\ee
Note that $\int d^2k_t\,f_{q/p}(x,k_t)=f_{q/p}(x)$. 
  
The inclusion of intrinsic $\vec{k}_t$-dependence in the parton
distributions is not straightforward from a theoretical point of view.
First of all, factorization theorems involving transverse momentum
dependence are generally harder to derive. Considerable
progress has been made for specific processes~\cite{CS-81} which are,
however, less complicated than the one considered here. We shall view
the factorized formulas employed below as reasonable phenomenological
extensions of collinear factorization at the semi-hard observed
transverse momenta we are considering. Furthermore, we refer to
ref.~\cite{Adrian1} for a derivation of collinear factorization from a
one-loop analysis of unpolarized parton scattering off a CGC target.

Another important theoretical issue regarding Sivers functions is the
calculable process dependence~\cite{Collins-02}. Since the
fitted functions from refs.~\cite{AM98,ABM-99} apply to the same
process as considered here (and to the same partonic subprocess in the
sense that $qg$ scattering is dominant in the considered kinematic
region), we will not worry about this issue here.

As already mentioned above, the Sivers functions appear in the
difference between upward and downward spin polarizations of the
proton 
\be 
d\sigma(p^\uparrow p\rightarrow hX) - d\sigma(p^\downarrow
p\rightarrow hX) &\propto& \int d^2k_t
[f_{q/p^\uparrow}(x,\vec{k}_t)-f_{q/p^\downarrow}(x,\vec{k}_t)]
\otimes d\sigma^{qp\rightarrow q'X}(\vec{k}_t) \otimes D_{h/q'}(z_h)
\nonumber\\ & = & \int d^2k_t \Delta^N f_{q/p^\uparrow}(x,\vec{k}_t)
\otimes d\sigma^{qp\rightarrow q'X}(\vec{k}_t) \otimes D_{h/q'}(z_h) ~,
\ee 
where $d\sigma^{qp\rightarrow q'X}(\vec{k}_t)$ denotes the cross
section of a parton scattering off an unpolarized proton. The fragmentation 
functions $D_{h/q'}(z_h)$ describe the fragmentation of a final-state
parton to a hadron with momentum fraction $z_h$. Note that the Sivers
functions correspond to distributions of unpolarized partons and so
the elementary hard-scattering cross sections are the usual
expressions for unpolarized partons, in both the numerator and the
denominator of $A_N$.

Because the convolution integrand has to be an even function of 
$\vec{k}_t$, while the Sivers function is odd, one finds
\be
\lefteqn{\int d^2k_t \Delta^N f_{q/p^\uparrow}(x,\vec{k}_t)
\otimes d\sigma^{qp\rightarrow q'X}(\vec{k}_t)}  
\nonumber\\
& = & \left(\int_{\hat{\vec{x}} \cdot \vec{k}_t> 0}
           +\int_{\hat{\vec{x}} \cdot \vec{k}_t< 0}\right)
d^2k_t\ \Delta^N f_{q/p^\uparrow}(x,\vec{k}_t) 
\otimes d\sigma^{qp\rightarrow q'X}(\vec{k}_t)
\nonumber\\
&=& \int_\mathrm{h.p.} d^2k_t\ \Delta^N f_{q/p^\uparrow}(x,\vec{k}_t) 
\otimes [d\sigma^{qp\rightarrow q'X}(\vec{k}_t)-
d\sigma^{qp\rightarrow q'X}(-\vec{k}_t)],
\label{eqn:kt_int}
\ee
where ``h.p.'' indicates that the integration extends over only the positive 
half plane with $\hat{\vec{x}} \cdot \vec{k}_t>0$. 

Following~\cite{Anselmino,AM98} we assume that the Sivers function
$\Delta^N f_{q/p^\uparrow}(x,\vec{k}_t)$ is sharply peaked about an
average transverse momentum $k_t^0=\sqrt{\langle\vec{k}_t^{\,2}\rangle}$
such that
\be
\Delta^N f_{q/p^\uparrow}(x,\vec{k}_t) \simeq 
\Delta^N f_{q/p^\uparrow}(x)\;\delta^{(2)}(\vec{k}_t-k_t^0\hat{\vec{k}}_t)~.
\label{eqn:difference2}
\ee
The unit vector $\hat{\vec{k}}_t$ is taken to lie
in the positive half of the scattering plane,
$\hat{\vec{x}}\cdot\hat{\vec{k}}_t>0$. Then
\be
\lefteqn{\int d^2k_t \; \Delta^N f_{q/p^\uparrow}(x,\vec{k}_t)
\otimes d\sigma^{qp\rightarrow q'X}(\vec{k}_t) \otimes D_{h/q'}(z_h)}
\nonumber\\
&\simeq&  
\Delta^N f_{q/p^\uparrow}(x) 
\otimes [d\sigma^{qp\rightarrow q'X}(k_t^0\hat{k}_t)-
d\sigma^{qp\rightarrow q'X}(-k_t^0\hat{k}_t)]\otimes D_{h/q'}(z_h).
\label{eqn:kt_int2}
\ee

Before the STAR data was published, the only available quantitative
knowledge of the Sivers functions $\Delta^N f_{q/p^\uparrow}(x)$ was
due to fits to SSA data from fixed-target $pp$ experiments and to low
energy semi-inclusive DIS data, cf.\ ref.~\cite{SiversInSidis} and
references therein. Although we will briefly comment on some more
recent fits below, for this paper we will mainly restrict ourselves to
the simpler parameterizations that resulted from fixed-target $pp$
data.  We only allow for a different overall normalization of the
Sivers functions. For valence quarks, the functions were parameterized
as follows~\cite{AM98,Boglione:1999pz}\footnote{We note that these
expressions are understood to apply at a fixed (average) scale. The
scale dependence is not known yet and will not be considered here.}:
\be
\Delta^N f_{q/p^\uparrow}(x) = K_{\Delta f}\,
                    \frac{k_t^0(x)}{M_p} \, N_q \, x^{a_q}\, (1-x)^{b_q}~.
\label{eqn:para_sivers}
\ee
The $x$-dependent average transverse momentum $k_t^0$
of valence quarks in the polarized proton is taken as
\be
k_t^0(x) = 0.47 \, x^{0.68} \, (1-x)^{0.48}\, M_p~,
\label{eqn:para_transv}
\ee
for all flavors. $M_p=1$~GeV denotes the proton mass. The maximal
intrinsic transverse momentum of $\approx210$~MeV is on the order 
of the inverse proton radius and therefore reasonable. At large $x$, 
one does not expect much higher intrinsic transverse momenta. The remaining
parameters are given by
\be
N_{u,d}/4=(3.68,-1.24),~~~a_{u,d}=
       (1.34,0.76),~~~b_{u,d}=(3.58,4.14)~. 
\label{eqn:sivers_pmset}
\ee
It was pointed out in ref.~\cite{ABM-99} that the original values of
$N_q$ from~\cite{AM98} were misquoted and have to be multiplied by a
factor of 4. In addition, in~(\ref{eqn:para_sivers}) we allow for
further adjustment of the magnitude of the Sivers functions by a
factor $K_{\Delta f}$ if required by the new high-energy data from
STAR.

\section{Forward particle production at high energies}

\subsection{Unpolarized $pp$ collisions}

The forward pion spectra in $pp$ collisions obtained by the STAR
Collaboration at RHIC ($\sqrt{s}=200$ GeV)~\cite{dAu_PRL} are
generally known to agree reasonably well with leading-twist NLO pQCD
calculations in the DGLAP approximation~\cite{Soffer,Guzey04}.
Although this approach is recognized as providing a much better
description of semi-hard particle production at RHIC than at lower
energy~\cite{Soffer}, not all commonly employed sets of parton
distribution and fragmentation functions (FFs) perform equally well.
To be specific, the NLO pQCD result with CTEQ6M PDFs and KKP FFs is
consistent with the data for rapidity $\langle\eta\rangle=3.8$, but
compares less well at $\langle \eta \rangle=4.0$, except in the
high-$p_t$ region. The Kretzer FFs lead to better agreement with
lower-$p_t$ data, presumably due to its smaller gluon fragmentation to
pions as compared to the KKP set, which may not be realistic though and 
one needs more precise extractions of the pion FFs at moderately high-$p_t$
($p_t=1\sim 3$~GeV/c) before definite conclusions can be drawn. The
fact that some standard PDFs and FFs fail to give a good description
at forward rapidities may find a natural explanation in the
observation that these data are sensitive to small-$x$ effects in the
target.

Hence, we would like to analyze these high-energy, forward
processes at RHIC within another theoretical approach, namely the
``Color Glass Condensate'' (CGC) formalism. Contrary to the DGLAP
approach, it resums small-$x$ effects in the gluon
distribution function. The Sivers functions, on the other hand, are
unaffected as the SSA arises from large-$x$ valence quarks on the
projectile side.

The CGC approach is most commonly applied to the study of gluon
saturation effects at sufficiently small $x$ and/or
$Q^2$~\cite{JK,nonlin}. At a given $x$, the transverse momentum below
which the unintegrated gluon distribution saturates due to non-linear
terms in the QCD evolution equations is called the saturation momentum
$Q_s(x)$. It grows approximately as a power of
energy~\cite{AM_DNT,Qsxgrowth}; NLO BFKL evolution gives $Q_s(x) \sim
1/x^\lambda$ with
$\lambda\simeq0.3$~\cite{Triantafyllopoulos:2002nz}. HERA
phenomenology indicates that $Q_s\sim1$~GeV at $x \sim
10^{-4}$~\cite{GBW}.

The deep saturation (non-linear) regime is therefore of minor
relevance to semi-hard ($p_t\ \gsim1$~GeV) particle production in $pp$
collisions at RHIC energy. This will be verified in a more
quantitative way in section~\ref{sec_HT}.  However, the CGC formalism
is also applicable to the so-called ``extended'' geometric scaling
(EGS) regime~\cite{AM_DNT,ScalingViol} which emerges above the saturation
line up to transverse momenta of about $Q_{gs}=Q_s^2/\Lambda\gg Q_s$,
where $\Lambda$ is a nonperturbative soft scale expected to be of
order $\Lambda_{\rm QCD}$. In deep inelastic scattering (DIS)
geometric scaling implies that at low $x \ (\lsim0.01)$ the proton
structure function depends only on the scaling variable
$\tau=Q^2/Q_s^2(x)$ rather than on $Q^2$ and $x$ separately.

Fits to DIS data from HERA~\cite{Stasto:2000er,IIM} found that
approximate geometric scaling holds not only at small $Q^2$, where
this is expected due to saturation effects, but over a much broader
window $Q_s^2 < Q^2 < Q_{gs}^2$, thereby providing evidence for 
the existence of such a new regime. It arises if the solution of the 
LO-BFKL evolution equation is expanded to second order around the saturation
saddle point~\cite{IIM}. The first term gives geometric scaling while
the second ``diffusion'' term contributes to scaling
violations~\footnote{To find $Q_{gs}\sim Q_s^2/\Lambda$ one should in
fact estimate $Q_{gs}$ from the transition point between the LLA and
DLA saddle points, respectively, rather than from the diffusion term
in the expansion of LO-BFKL about the saturation saddle point. For
details see, for example, section~2.4.3 in ref.~\cite{JK}
and references therein.}. 
Due to this diffusion term, the anomalous
dimension $\gamma$ governing the small-$x$ evolution of the gluon
distribution with rapidity $y=\log(1/x)$ is shifted from
$\gamma_s\simeq 0.63$ (the BFKL saddle point in the vicinity of the
saturation line) by $\Delta\gamma\propto \log(1/r_tQ_s)/y$, where
$r_t\sim1/Q$ denotes the dipole size. As $Q$ approaches $\sim Q_{gs}$
from below, violations of geometric scaling grow to order unity and
the anomalous dimension reaches $\gamma_{\rm
DGLAP}\sim1$ (more precisely, $1-\gamma_{\rm DGLAP}= {\cal
O}(\alpha_s)$). Ref.~\cite{IIM} finds that the average shift required
for a good fit to the HERA data is not small,
$\langle\Delta\gamma\rangle\simeq 0.2$.  Thus, the behavior of the
anomalous dimension which determines the (geometric) scaling
violations plays a key role for high-energy $ep$ scattering.

\begin{figure}[htb]
\centering
\centerline{\epsfig{figure=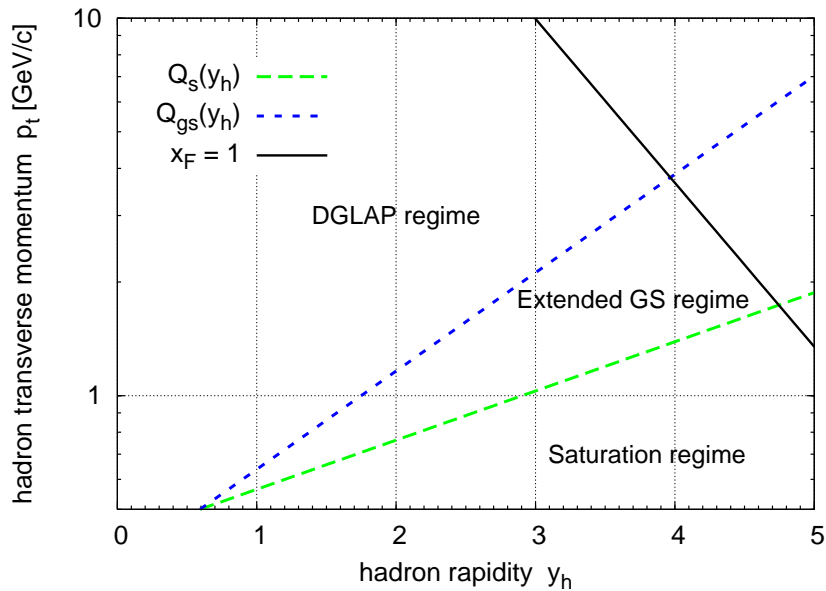,width=4.5in}}
\caption{``Phase diagram'' of the target proton at RHIC energy
($\sqrt{s}=200$ GeV).  $y_h$ and $p_t$ are the rapidity and the
transverse momentum of the produced hadron, respectively.  The dashed
line depicts the boundary $Q_s(y_h)$ of the saturation region;
the dotted line that of extended geometric scaling, $Q_{gs}(y_h)$. For
this plot we fixed the projectile parton's momentum fraction to the
typical value $x_1=0.1$. The solid line corresponds to the
boundary of phase space, $x_F= 1$.}
\label{fig:QCDphase}
\end{figure}
The same approach has also been applied to high-$p_t$ ($\gsim1$~GeV)
hadron production in $d\, Au$ collisions at RHIC over a wide interval
of rapidity $y_h$~\cite{Adrian2}. Indeed, the data was shown to agree
rather well with a parameterization of the anomalous dimension which
increases logarithmically with $p_t$ from $\gamma=\gamma_s$ to its
asymptotic value $\gamma\sim1$, while decreasing with $y$ 
as $\Delta\gamma\sim1/y$
at very large rapidity. The $d\, Au$ data from RHIC is therefore
consistent with the presence of an extended geometric scaling regime
above the saturation region which widens towards forward rapidities
(i.e.\ smaller $x_2$). The EGS window is characterized by the need to
resum logarithms of $1/x_2$ rather than of $Q^2$. However, the twist
expansion remains valid, all-twist resummation being necessary only in
the saturation regime $Q^2\lsim Q_s^2$.

We illustrate the various kinematic regimes emerging in $pp$
collisions at RHIC in Fig.~\ref{fig:QCDphase}. The boundaries of
these regions were determined as follows. The saturation regime of
the target proton extends to 
\be  \label{Qsx2}
Q_s(x_2) = 1\,\mathrm{GeV}\,\left(\frac{x_0}{x_2}\right)^{\lambda/2}~,
\ee
where, as already referred to above, $x_0=3.0\times 10^{-4}$ and
$\lambda=0.3$~\cite{GBW}. $x_2$ is related to the rapidity $y_h$ of
the produced hadron by kinematics: $x_1/x_2=\exp(2y_h)$ (see, for
example, appendix~B in~\cite{Adrian1}). For this figure, we fixed
$x_1$ to a value typical for a valence quark, $x_1=10^{-1}$.  The EGS
regime corresponds to the window between $Q_s$ and $Q_{gs}$, where the
anomalous dimension $\gamma$ (given below in eq.~(\ref{eq:gam_new})) grows 
from $\gamma_s$ to $\sim1$. In
fig.~\ref{fig:QCDphase}, its upper limit is taken to be $Q_{gs} =
Q_s^2 /\Lambda$ with $\Lambda = 0.5$~GeV, for illustration. Above $Q_{gs}$,
$\gamma\sim\gamma_{\rm DGLAP}$ and so the DGLAP approximation applies;
note that this of course includes the DLL regime of large $y_h$ and
large $p_t$. The boundary of phase space, finally, is reached when the
Feynman-$x$ of the produced hadron $x_F=(p_t/\sqrt{s})\exp(y_h) = 1$.

This figure confirms that the {\em all-twists} saturation regime plays
no prominent role at RHIC energy and $p_t\, \gsim1$ - $1.5$~GeV/c,
regardless of rapidity. However, for $y_h\, \gsim3$ the EGS window opens
wide and eventually covers essentially the entire semi-hard regime of
particle production as the DGLAP region is cut off by energy-momentum
conservation constraints. This implies that at such rapidities, $\gamma$
does not reach 1 at any accessible transverse momentum. It may
therefore be plausible that small-$x$ effects are relevant for
semi-hard particle production in the forward region of $pp$
collisions at RHIC.

Forward particle production leads to a strong asymmetry in the
momentum fractions $x_1\sim1$ and $x_2\ll1$ which are probed in the projectile
and target, respectively. The process can then be described using an
asymmetric DGLAP $\otimes$ CGC factorization scheme developed in
ref.~\cite{Adrian1},
\be
{dN_h \over dy_h d^2p_t} &=& 
{1 \over (2\pi)^2} \int_{x_F}^{1} dx_1 \, {x_1\over x_F}
\Bigg[f_{q/p}(x_1,p_t^2)\, N_F \left({x_1\over x_F}p_t,x_2\right)\,
D_{h/q}\left({x_F\over x_1},p_t^2\right)
\nonumber \\
& &+~
f_{g/p}(x_1,p_t^2)\, N_A \left({x_1\over x_F}p_t,x_2\right)\, 
D_{h/g}\left({x_F\over x_1},p_t^2\right)\Bigg]~.
\label{eq:conv2}
\ee 
The indices $q$ in $f_{q/p}$ and $D_{h/q}$ are implicitly summed over
all participating quark flavors. In what follows, the mass of the
produced hadrons is neglected because $m_\pi \ll p_t$, which allows us
to equate to a good approximation the pseudorapidity $\eta$ with the
rapidity $y_h$.

The approach~(\ref{eq:conv2}) includes DGLAP evolution of the
projectile parton distribution functions $f_{q/p}(x_1,p_t^2)$ and of
the fragmentation functions $D_{h/q}(x_F/x_1,p_t^2)$ with the
factorization scale $Q^2=p_t^2$. It resums small-$x$ logarithms in the
target as well as DGLAP logarithms of $Q^2$ on the projectile side and
has been derived to leading order in $\alpha_s$~\cite{Adrian1}.  Exact
rather than small-$x$ approximated splitting functions are to be
employed, such that the projectile momentum is conserved, which is
important at large $x_1$. On the other hand, the target proton is
described by the dipole forward scattering amplitudes $N_{F,A}$, where
$N_F$ corresponds to a projectile quark impinging on the target's
small-$x$ gluons, while $N_A$ applies to a projectile gluon. These
arise~\cite{GelisJalilian} as averages of two-point functions of
Wilson lines in the corresponding representation (running up and down
the light-cone, separated by a transverse distance $r_t$) over the
gluon field of the hadron~\cite{Balitsky:1995ub}. In the high-energy
limit one therefore recovers Mueller's dipole
approach~\cite{MuellerDipole}. For a recent comparison of this
approach with HERA diffractive electroproduction data we refer to
ref.~\cite{Forshaw:2006np}.

The dipole profile employed here is similar to that from the fit
presented in ref.~\cite{GBW} to HERA DIS data, except for the presence
of the anomalous dimension $\gamma<1$~\cite{IIM,KKT,JK,Adrian1,Adrian2}:
\be 
N_A({r}_t,x_2) =
1-\exp\left[-\frac{1}{4}(r_t^2 Q_s^2(x_2))^{\gamma(r_t,x_2)}\right]~.
\label{NA_param}
\ee
${r}_t$ denotes the transverse size of the dipole. $N_F$ is obtained
here by the replacement $Q_s^2\to (C_F/C_A)\,Q_s^2=(4/9)\,Q_s^2$,
corresponding to a weaker coupling of a quark relative to a gluon
projectile to the target field. Due to this color factor, our
effective saturation momentum entering $N_F$ is somewhat smaller than
the original Golec-Biernat$-$W\"ustoff (GB-W) fit~\cite{GBW} but
agrees with previous comparisons to RHIC
data~\cite{Adrian1,Adrian2,KKT} and with the analysis of HERA DIS data
presented in ref.~\cite{IIM}.

The {\sl Ansatz}~(\ref{NA_param}) exhibits saturation at $r_t>1/Q_s$
and also reproduces the $N_A\sim r_t^2$ DGLAP limit as $r_t\to0$
(assuming that also $\gamma\to1$ in that limit)\footnote{Note that in
order to reproduce the $\sim1/q_t^4$ DGLAP power-law tail in {\em
momentum} space one needs to introduce an additional logarithm of
$1/r_t^2$ in the exponent of~(\ref{NA_param}). For $\gamma=1$,
$N_A(q_t)$ then drops approximately exponentially in the high-density
regime about $q_t\sim Q_s$, turning into $\sim 1/q_t^4$ LO DGLAP
behavior as $q_t\to\infty$~\cite{Boer:2002ij,IIT}. A similar
modification of the GB-W model was considered in
ref.~\cite{Bartels:2002cj}.}. It also displays exact geometric scaling
when $\gamma$ is constant since then $N(r_t,x)$ depends only on the
scaling variable $\rho\equiv r_t Q_s$. Violations of geometric scaling
arise due to the dependence of $\gamma$ on $r_t$, increasing from
$\gamma_s$ at $r_t\sim1/Q_s$ to $\gamma_{\rm DGLAP}\sim1$ at very
short distances. Here, the anomalous dimension $\gamma(r_t,x_2)$ of
the gluon distribution is parameterized as
\be 
\gamma(r_t,x_2) &=&
\gamma_s + (1-\gamma_s)\, \frac{\log(1/r_t^2Q_s^2(x_2))} {\lambda
y_2+d\sqrt{y_2}+\log(1/r_t^2Q_s^2(x_2))}
\label{eq:gam_new}
\ee 
with $\gamma_s\simeq 0.627$; $y_2=\log\,1/x_2$ is (minus) the rapidity
of the target partons. This function agrees with the solution of
LO-BFKL evolution with saturation boundary conditions in the saddle
point approximation~\cite{IIM}, and with the general theoretical
limits mentioned above. However, the
parameterization~(\ref{eq:gam_new}) also includes subleading
corrections $\sim d\sqrt{y_2}$ which govern geometric scaling
violations at subasymptotic rapidities. With $d\simeq1.2$ the
anomalous dimension~(\ref{eq:gam_new}) was shown to provide a good
description of $d\,Au$ data from RHIC over a wide range of
rapidity~\cite{Adrian2}. We note that~(\ref{eq:gam_new}) applies only
in the EGS regime, while $\gamma=\gamma_s$ stays fixed within the
saturation region (which is, however, not important here, see
section~\ref{sec_HT}).

\begin{figure}[hbt]
\centering
\centerline{\epsfig{figure=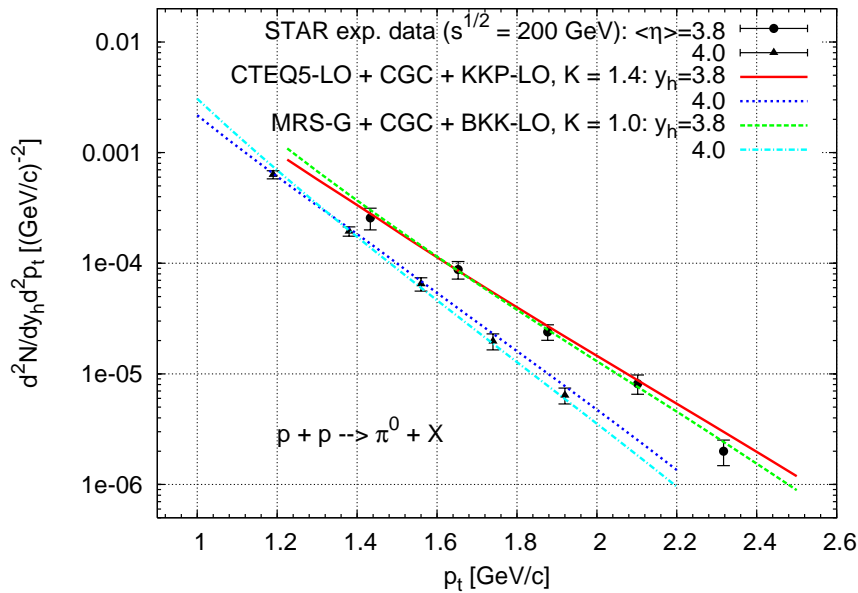,width=3.2in,angle=-90}}
\caption{Transverse momentum distributions of forward inclusive
$\pi^0$ from unpolarized $pp$ collisions at $\sqrt{s}=200$~GeV. The
STAR data~\protect\cite{dAu_PRL} were taken at average
pseudorapidities of $\langle \eta \rangle=3.8$ and $4.0$,
respectively. Theoretical results correspond to two different sets of
parton distribution and fragmentation functions, i.e.\ CTEQ5-LO+KKP-LO
(I) and MRS-G+BKK-LO (II), respectively, all evaluated at
$Q^2=p_t^2$. They give $\chi^2$ per data point of 0.9 (0.3) at $y=3.8$
and 0.7 (1.0) at $y=4.0$ for set I (II).}
\label{fig:pp_fwd3840_pi}
\end{figure}

To determine the transverse momentum distribution of hadrons via
eq.~(\ref{eq:conv2}) requires taking the Fourier transforms of the dipole
profiles~(\ref{NA_param}) to momentum space. To simplify this task, we
shall replace $\gamma(r_t,x_2)$ by $\gamma(1/q_t,x_2)$, where $q_t =
(x_1/x_F) p_t$ is the Fourier conjugate of $r_t$. This provides a
good approximation at large $y_h$ (small $x_2$) since there $\gamma$
increases rather slowly with $q_t$.

In fig.~\ref{fig:pp_fwd3840_pi} we present our numerical results for
the $p_t$-distributions of forward inclusive pions from $pp$
collisions at $\sqrt{s}=200$~GeV. They correspond to rapidities
$y_h=3.8$ and $4.0$, respectively.  We employ two different sets of
parton distribution (PDF) and fragmentation (FF) functions: CTEQ5-LO
PDFs~\cite{cteq} and KKP-LO FFs~\cite{KKP}, which is nowadays a common
combination. The other set is MRS-G PDFs~\cite{Martin:1995ws} and
BKK-LO FFs~\cite{Binnewies:1994ju} as used in ref.~\cite{AM98} to
extract the parameterization of the Sivers functions from experimental
data obtained at lower energies.  We observe rather good agreement of
both sets in the interval $1.2\lsim p_t \lsim 2.4$ GeV/c with recent
STAR data from RHIC~\cite{dAu_PRL}. For the overall
normalization, we need to multiply our LO results by a factor $K\simeq
1.4$ for the CTEQ5-LO+KKP-LO set I, while $K\simeq 1.0$ for the
MRS-G+BKK-LO set II. The $\chi^2$ per data point is 0.9
(0.3) at $y=3.8$ and 0.7 (1.0) at $y=4.0$ for set I (II), which is
rather good. Since both sets work equally well within error bars, we
will henceforth employ set II which does not require a K-factor.

The results from fig.~\ref{fig:pp_fwd3840_pi} correspond to a scale
$Q^2=p_t^2$. The scale dependence in eq.~(\ref{eq:conv2}) arises
entirely from the LO PDFs and FFs and amounts to a weakly
$p_t$-dependent normalization factor~\cite{Adrian1}. If the scale
dependence of the LO Sivers functions turns out to be similar to that of the
unpolarized PDFs, such an overall $K$-factor would cancel in the ratio
of the polarized to unpolarized processes. The evolution properties of
the Sivers functions are, however, largely unknown at present and as a
consequence we have to postpone a more detailed study of such issues
to the future.

\subsection{$p^\uparrow p$ collisions 
and single transverse-spin asymmetry}

The single-inclusive pion distribution of eq.~(\ref{eq:conv2}) is
written in a collinearly factorized convolution form, $d\sigma(p
p\rightarrow hX) \sim f_{q/p} \otimes N_F \otimes D_{h/q}$. We shall
formulate the single transverse-spin asymmetry $A_N$ in an analogous
way but partly include transverse momentum dependence in order to
incorporate the Sivers effect. For the numerator of
eq.~(\ref{eqn:asymmetry}) this results in
\be
d\sigma(p^\uparrow p\rightarrow hX) - d\sigma(p^\downarrow p\rightarrow hX)
&\propto& \int d^2k_t
[f_{q/p^\uparrow}(x_1,\vec{k}_t)-f_{q/p^\downarrow}(x_1,\vec{k}_t)]
\otimes N_F(x_1,\vec{q}_t,\vec{k}_t) \otimes D_{h/q}(x_F/x_1)
\nonumber\\
&\simeq& \Delta^N f_{q/p^\uparrow}(x_1) \otimes
[N_F(x_1,q_t-k_t^0)-N_F(x_1,q_t+k_t^0)] \otimes D_{h/q}(x_F/x_1)~,
\label{eqn:an1}
\ee 
where we used eq.~(\ref{eqn:kt_int}) to arrive at the last line.
Also, we simplified the kinematics (as was done in
refs.~\cite{AM98} and therefore has to be considered as part of the
fit of the Sivers function) by picking up only contributions where
$\vec{k}_t$ is either parallel or anti-parallel to $\vec{q}_t$.

On the other hand, the denominator of~(\ref{eqn:asymmetry}) is given by
\be
d\sigma(p^\uparrow p\rightarrow hX) + d\sigma(p^\downarrow p\rightarrow hX)
&\propto& \int d^2k_t
[f_{q/p^\uparrow}(x_1,\vec{k}_t)+f_{q/p^\downarrow}(x_1,\vec{k}_t)]
\otimes N_F(x_1,\vec{q}_t,\vec{k}_t) \otimes D_{h/q}(x_F/x_1)
\nonumber\\
&\simeq& 2 f_{q/p}(x_1) \otimes N_F(x_1,q_t) \otimes D_{h/q}(x_F/x_1)~,
\label{eqn:an2}
\ee
where eq.~(\ref{eqn:sum}) and the fact that $f_{q/p}(x,\vec{k}_t)$ is
an even function of $\vec{k}_t$ have been used.
Thus, when displaying explicitly all arguments of the various
functions the asymmetry becomes
\be
A_N(p_t,y_h) &=& \frac{1}{2}
{\sum\limits_{{\rm val}-q} \Delta^N f_{q/p^\uparrow}(x_1) \otimes
\left[N_F\left({x_1\over x_F}p_t-k_t^0,y_2\right)
-N_F\left({x_1\over x_F}p_t+k_t^0,y_2\right)\right] 
\otimes D_{h/q}\left({x_F\over x_1},p_t^2\right)
\over
\sum\limits_q f_{q/p}(x_1,p_t^2) \otimes N_F\left({x_1\over x_F}p_t,y_2\right)
\otimes D_{h/q}\left({x_F\over x_1},p_t^2\right)
+ f_{g/p}(x_1,p_t^2) \otimes N_A\left({x_1\over x_F}p_t,y_2\right)
\otimes D_{h/g}\left({x_F\over x_1},p_t^2\right)}.
\nonumber\\
\label{eqn:an3}
\ee
Only valence-quark contributions are accounted for in the numerator,
according to the parameterization of the Sivers functions from
ref.~\cite{AM98}. On the other hand, the unpolarized cross section in
the denominator includes the contributions of all active quark flavors
and of gluons. At large rapidities, however, it is also dominated by
valence quarks. It is evident from~(\ref{eqn:an3}) that a good
description of the elementary quark $q_t$ distribution and of its
{\em derivative} is required to reliably extract the Sivers function
from $A_N$.

\begin{figure}[ht]
\centering
\centerline{\epsfig{figure=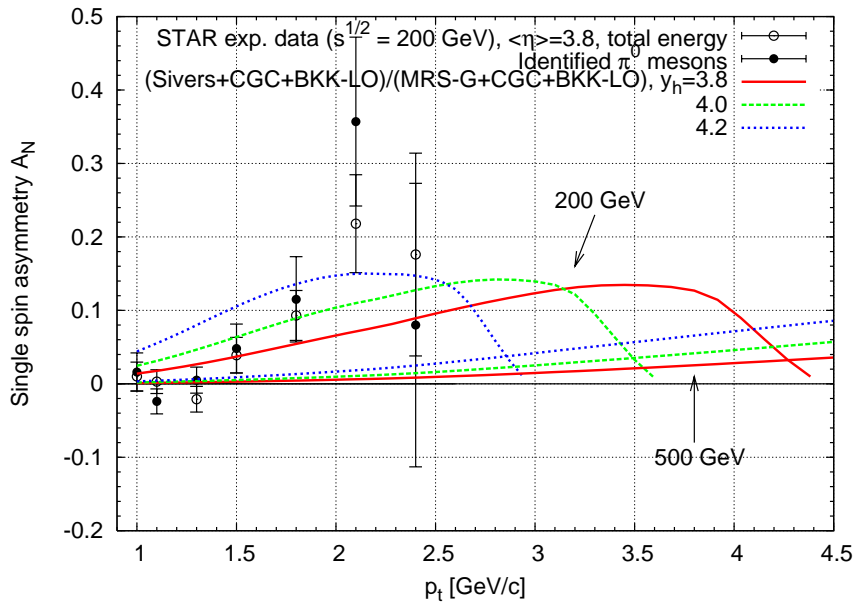,width=3.2in,angle=-90}}
\caption{Single transverse-spin asymmetry $A_N$ in the rapidity
interval $y_h=3.8$ - 4.2 for $\sqrt{s}=200$, 500~GeV, with $K_{\Delta
f}=2$ (two times larger Sivers functions than ref.~\cite{AM98}). The
STAR data~\cite{STAR03} were taken at an average pseudorapidity of
$\langle\eta\rangle=3.8$ and $\sqrt{s}=200$~GeV.}
\label{fig:AN_pp_fwd3840_pi}
\end{figure}
In Fig.~\ref{fig:AN_pp_fwd3840_pi} we compare our numerical results
for the asymmetry $A_N$ to STAR data~\cite{STAR03} taken at
$\sqrt{s}=200$~GeV. We also show predictions for larger rapidities,
$y_h=4.0$ and 4.2, and for higher energy, $\sqrt{s}=500$~GeV. The
theoretical curves were obtained from eq.~(\ref{eqn:an3}) with the
MRS-G~\cite{Martin:1995ws} PDFs for the unpolarized projectile proton
and the BKK-LO~\cite{Binnewies:1994ju} FFs, the
dipole~(\ref{NA_param}), the anomalous dimension~(\ref{eq:gam_new}),
and the saturation momentum~(\ref{Qsx2}). For the Sivers functions, we
adopted the parameterization
(\ref{eqn:para_sivers},\ref{eqn:para_transv},\ref{eqn:sivers_pmset})
with $K_{\Delta f}=2$, corresponding to a two times larger Sivers
effect as compared to the original fit from ref.~\cite{AM98}.

One observes that our curve at $y_h=3.8$ goes up more slowly than the
data and peaks at large $p_t$ around $p_t\sim3.5$~GeV/c. At higher
rapidity, however, the curves increase more steeply and the peak
shifts to smaller $p_t$. At fixed rapidity, the asymmetry decreases
rather rapidly with energy. This is mainly a kinematic effect in that
the typical momentum {\em fraction} $x_1$ of projectile partons
decreases, and so does the Sivers function~(\ref{eqn:para_sivers}).
Given the theoretical uncertainty in the Sivers functions and the
sizable errors of the data, the agreement can be considered
reasonable. Even with the present large error bars, however, the STAR
data appear to imply larger Sivers functions than previously
thought~\cite{AM98} (as reflected by $K_{\Delta f}=2$), at least if
$A_N$ indeed arises entirely from the Sivers effect. If more precise
data becomes available in the future, our formalism (extended to allow
for more general $\vec{k}_t$ dependence of the Sivers functions if
needed) could be used for a complete refit of the Sivers functions.

A few additional comments regarding the factor $K_{\Delta f}\, \gsim2$
found here are in order. The
parameterization~(\ref{eqn:para_sivers}-\ref{eqn:sivers_pmset}) of the
Sivers function resulted from a fit to low-energy fixed-target data,
for which an accurate description of the pion transverse momentum
distribution (and of its slope) is hard to obtain. It is
therefore not entirely surprising that the new high-energy data, while
confirming qualitative features of the parameterizations, may still
require some quantitative adjustments. Of course, such adjustments 
would alter predictions for asymmetries of other processes, like for instance
the Drell-Yan process~\cite{ADM}. When high energy data with smaller error
bars becomes available we intend to perform a more detailed
investigation of the $x$- and $\vec{k}_t$-dependence of the Sivers functions, 
rather than of their overall normalization only.

We should also point out that a good description of the STAR data
can also be achieved by increasing the intrinsic transverse
momentum $k_t(x)$ of valence quarks substantially, 
as recent parameterizations show~\cite{DM04}. This is because the Sivers
distribution~(\ref{eqn:para_sivers}) is proportional to the intrinsic
transverse momentum, and another factor of $k_t$ arises in eq.~(\ref{eqn:an1})
from the difference of the contributions of quarks with $\vec{k}_t$
parallel or anti-parallel to the momentum transfer; see for example
eq.~(\ref{Delta_qXsec}) below. However, the process
considered here is dominated by large $x$ on the
projectile side ($x_1\sim 0.1$ - 1), where intrinsic transverse momenta 
should probably not exceed the inverse proton radius by much. This is
consistent also with our finding (and that from NLO DGLAP approaches)
that the forward inclusive pion distribution from unpolarized
collisions can be described rather well without assuming large
intrinsic transverse momenta of projectile valence quarks.

\section{Higher-twist effects on SSA} \label{sec_HT}
The unitarized dipole profiles $N_{F,A}$
entering~(\ref{eqn:an1},\ref{eqn:an2}) resum all higher-twist effects
which arise at low transverse momentum. It is interesting to
disentangle the leading-twist contribution to the observables
considered here. For $p_t>Q_s$, one expects that higher-twist
contributions are suppressed.

We estimate the leading contribution by introducing
\be 
N_F(r_t,y_2;c) & \equiv & \frac{1}{c}
\left\{1-\exp\left[-\frac{c}{4}(4 r_t^2
Q_s^2(y_2)/9)^{\gamma(r_t,y_2)} \right]\right\}~.  \label{NFc} \ee
For $c=1$ we recover the full $N_F(r_t,y_2)$ from~(\ref{NA_param}).
In the limit $c\rightarrow 0$, however, the Fourier transform
of~(\ref{NFc}) isolates the leading contribution in the expansion with
respect to $Q_s/q_t$. For $q_t \neq 0$ one obtains
\be
N_F(q_t\neq 0,y_2;c) &=& - \int d^2r_t\ 
e^{i\vec{q}_t\cdot \vec{r}_t} N_F(r_t,y_2;c)
\nonumber\\
&=& \frac{2\pi}{c}\int_0^\infty dr_t\ r_t J_0(r_t q_t)
\,\exp\left[-\frac{c}{4}(4 
r_t^2 Q_s^2(y_2)/9)^{\gamma(r_t,y_2)}\right].
\label{twist1}
\ee
We know analytic forms of the Fourier transform~(\ref{twist1})
only for $\gamma=1/2$ and 1~\cite{Adrian1}.
In the limit $c\rightarrow 0$ one finds
\be
N_F^{\gamma=1/2}(q_t,y_2;c) 
&=& \frac{\pi Q_s}
{3 \left[\left(Q_s c/6\right)^2+q_t^2\right]^{3/2}}
\stackrel{c\rightarrow 0}{\longrightarrow} \frac{\pi}{3Q_s^2} \left({Q_s\over
  q_t}\right)^3~,  \label{NF_LT_BFKL}
\\
N_F^{\gamma=1}(q_t,y_2;c) 
&=& \frac{9\pi}{Q_s^2c^2} \exp\left[-\frac{9}{4c}\left(\frac{q_t}{Q_s}
\right)^2\right]
\stackrel{c\rightarrow 0}{\longrightarrow} 0~.
\label{twist2}
\ee
Hence, for $\gamma=1/2$, representing the LO-BFKL anomalous dimension
{\em without} saturation boundary conditions, the Fourier transformed
dipole profile exhibits a leading-twist tail $\sim
(Q_s/q_t)^3$ (we recall that $\sim (Q_s/q_t)^4$ emerges in
the DGLAP regime). On the other hand, the GB-W model~\cite{GBW}
corresponding to $\gamma=1$ shows no power-law tail at large $q_t$. It
arises entirely from all-twist resummation and has no expansion
in powers of $Q_s/q_t$.

\begin{figure}[hbt]
\centering
\centerline{\epsfig{figure=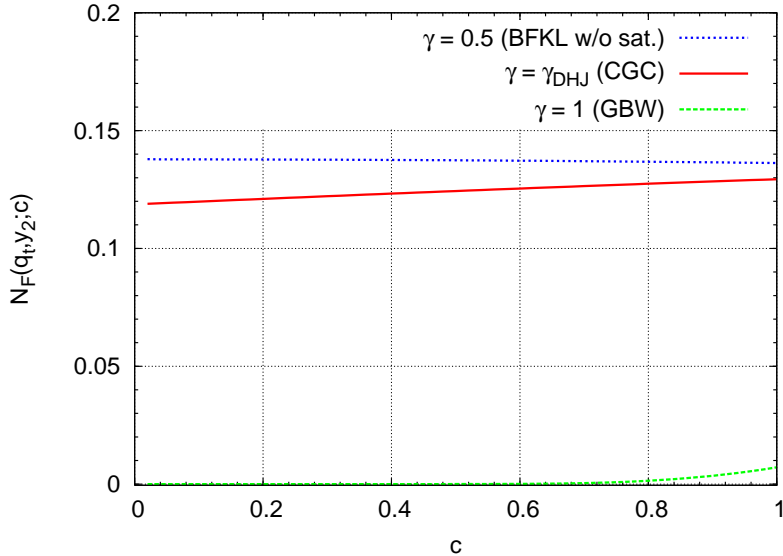,width=3.0in,angle=-90}}
\caption{Study of leading and higher twist contributions in the dipole 
profile $N_F(q_t,y_2)$.
We plot the results of eq.~(\ref{twist1}) with $\gamma=0.5$, 1 and 
$\gamma_\mathrm{DHJ}$ from~(\ref{eq:gam_new}), respectively, as a
function of $c$ at $y_h=3.8$, $p_t=2$~GeV/c and $x_1=x_F$.}
\label{fig:AN_LTcont_unpol}
\end{figure}
For the anomalous dimension from~(\ref{eq:gam_new}) the Fourier
transform and its $c \to 0$ limit cannot be obtained analytically. We
therefore show numerical results in Fig.~\ref{fig:AN_LTcont_unpol}.
Here, we set $y_h=3.8$, $p_t=2$~GeV/c, which represent typical values,
and take $x_1=x_F$ such that $p_t=q_t$.

One observes that the profile for $\gamma=1/2$ is almost constant in
the interval $0\le c \leq 1$ and approaches a finite value as $c
\rightarrow 0$, which agrees with the analytic
result~(\ref{NF_LT_BFKL}). This confirms that for these kinematic
conditions the dipole with anomalous dimension from LO-BFKL without
saturation boundary condition is dominated entirely by the
leading-twist contribution.  Next, we consider the $q_t$-dependent
anomalous dimension~(\ref{eq:gam_new}) which generates violations of
geometric scaling and interpolates to the DGLAP regime at high
$q_t$. For this case, the Fourier transformed dipole profile decreases
slowly as $c\to 0$ from its value at $c=1$. This decrease corresponds
to higher-twist contributions of about $8\%$.  Thus, higher-twist
contributions to single-inclusive forward hadron production in $pp$
collisions are not very large once an anomalous dimension with the
described features is taken into account.  Quantitative comparisons to
the anomalous dimension from leading-twist NLO DGLAP approaches for
the present kinematics would be interesting.

\begin{figure}[hbt]
\centering
\centerline{\epsfig{figure=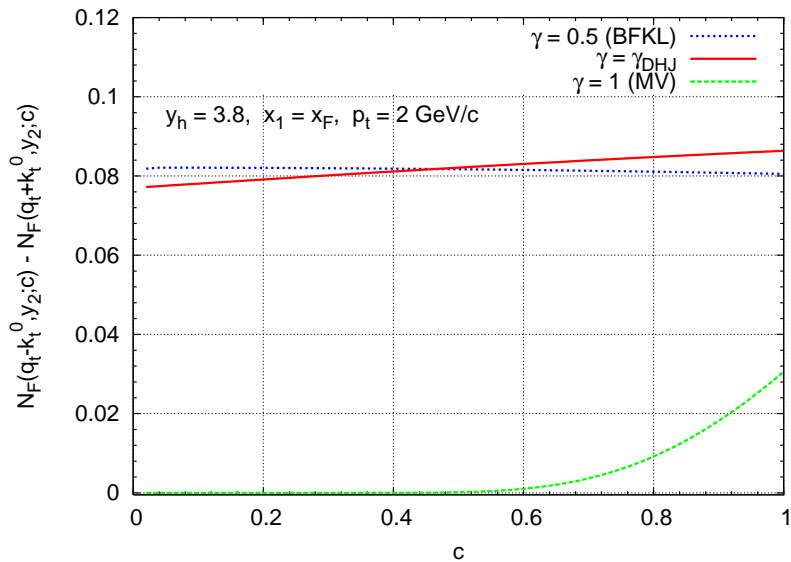,width=3.0in,angle=-90}}
\caption{Same as Fig.~\ref{fig:AN_LTcont_unpol}
but now for the difference $N_F(q_t-k_t^0)-N_F(q_t+k_t^0)$.}
\label{fig:AN_LTcont_sivers}
\end{figure}
Next, we address the same question for the difference appearing in the
numerator of the asymmetry $A_N$ which essentially corresponds to the
derivative of $N_F(q_t)$. For this quantity, the leading contribution
is suppressed by an additional power of $k^0_t/q_t$ as can be easily
realized by considering the leading contribution for $\gamma=1/2$ as
an example:
\be
\frac{Q_s^3}{(q_t-k^0_t)^3} - \frac{Q_s^3}{(q_t+k^0_t)^3}
= 6 \frac{Q_s^3}{q_t^3}\frac{k^0_t}{q_t}+\cdots    \label{Delta_qXsec}
\ee
Due to this additional factor, the magnitude of
$N_F(q_t-k_t^0)-N_F(q_t+k_t^0)$ as shown in
fig.~\ref{fig:AN_LTcont_sivers} is smaller than that of $N_F(q_t)$
from fig.~\ref{fig:AN_LTcont_unpol}. Here, too, the $c\to0$ limit isolates
the leading power of $Q_s/q_t$ from this quantity.
The curve for $\gamma=1/2$ is again flat over the entire interval, 
implying leading-twist dominance. On the other hand, for
$\gamma=1$ (GB-W) the leading-twist contribution vanishes completely.
For the $q_t$-dependent anomalous dimension~(\ref{eq:gam_new}) we observe
higher-twist contributions of about $11\%$.

\begin{figure}[hbt]
\centering
\centerline{\epsfig{figure=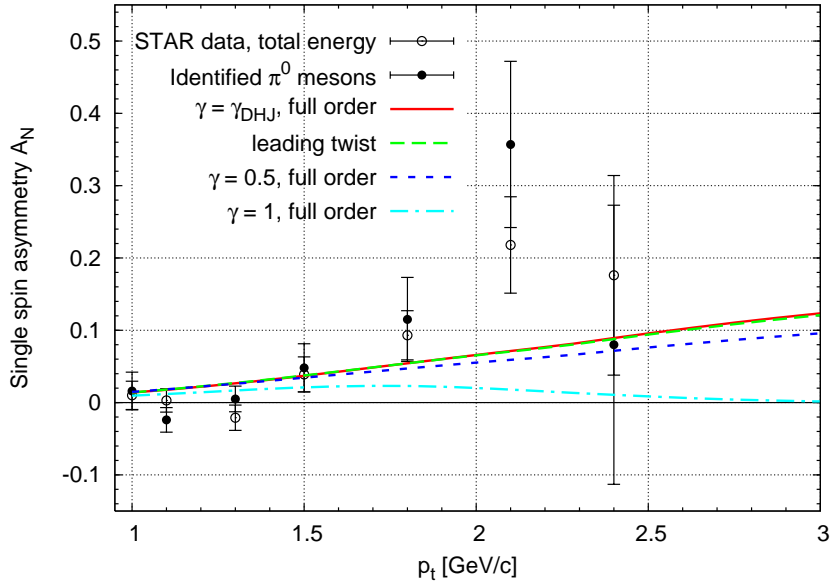,width=4.5in}}
\caption{The influence of the anomalous dimension on the asymmetry
$A_N$ at $y_h=3.8$ and $1\leq p_t \leq 3$~GeV/c.}
\label{fig:AN_ppf3840_pi0_STAR6}
\end{figure}
Finally, Fig.~\ref{fig:AN_ppf3840_pi0_STAR6} depicts the asymmetry
$A_N$ over the range $1\leq p_t \leq 3$ GeV/c at $y_h=3.8$ and
$\sqrt{s}=200$~GeV for $\gamma=1/2$, 1, and $\gamma=\gamma_{\rm DHJ}$
from eq.~(\ref{eq:gam_new}). For the latter case, we also show $A_N$
as obtained from the leading-twist contribution to $N_F$ alone. Since
the full and leading-twist curves are almost identical, we conclude
that higher-twist contributions in the numerator and denominator
cancel to a large extent. On the other hand, it turns out that the
growth of the anomalous dimension with transverse momentum (geometric
scaling violations) does give rise to a slightly steeper $A_N(p_t)$ as
compared to a constant $\gamma=1/2$. The GB-W dipole, which does not
permit a twist expansion, gives a very small asymmetry. The reason for
this behavior is the unrealistically large contribution from gluons to
the unpolarized cross section in the denominator of $A_N$ at high
rapidity: in momentum space, $N_A(q_t)\sim\exp[-q_t^2/Q_s^2]$ falls
off less rapidly with $q_t$ than
$N_F(q_t)\sim\exp[-(C_A/C_F)\,q_t^2/Q_s^2]$.

\section{Summary}

In this paper, we have analyzed recent high-energy data from STAR on
inclusive pion production at forward rapidities and on single
transverse spin asymmetries (SSA) in $p^\uparrow p$
collisions. Particle production at large rapidities involves
very small momentum fractions in the target, hence we propose and
use a factorization approach which accounts for small-$x$
effects on the gluon distribution of the target. Within this ``Color
Glass Condensate'' approach, the target is described by unitarized two-point
functions of light-like Wilson lines (dipole forward scattering
amplitudes) rather than the usual leading-twist, leading-log (DGLAP)
gluon distribution function.

We have shown explicitly that the deeply non-linear (``saturation'')
regime is of little relevance for particle production in $pp$
collisions at RHIC energy. The dipole scattering amplitude is
therefore well approximated by its leading-twist limit $N(r_t,x) \sim
(r_t Q_s(x))^{2\gamma(r_t,x)}$. Nevertheless, small-$x$ evolution not
only leads to saturation for transverse momenta $q_t\le Q_s(x)$ but
also to a so-called ``extended geometric scaling'' window {\em above}
$Q_s$, which has been confirmed by fits to HERA DIS data. Within the
EGS window extending from $Q_s(x)$ to $Q_{gs}(x)\sim Q_s^2(x) /
\Lambda \gg Q_s(x)$, the anomalous dimension $\gamma$ of the gluon
distribution function increases smoothly from its value at the BFKL
saddle-point with saturation boundary conditions,
$\gamma_s\simeq0.63$, to its value in the dilute DGLAP regime,
$\gamma_{\rm DGLAP}\sim1$. Once this feature is incorporated, a very
good baseline for forward pion production in unpolarized $pp$
collisions is obtained ($\chi^2$ per data point $\approx1$). A
comparison of our $\gamma$ to that from NLO DGLAP approaches in the
kinematic regime relevant for STAR would be interesting, given that
the predicted pion $p_t$-distributions are similar.

Building on this baseline, we then proceed to analyze single
transverse-spin asymmetries in the forward region. We restrict
ourselves to SSAs originating from the Sivers effect, which is
nowadays believed to give the dominant contribution within the
framework of transverse spin and transverse momentum dependent parton
distribution functions. The approximately $10\%$ contributions from
higher twists to the polarized and unpolarized cross sections,
respectively, largely cancel in their ratio $A_N$, which is therefore
very well approximated by the leading-twist contribution. Furthermore,
we find that a parameterization of the Sivers functions obtained
previously from a fit to fixed-target $pp$ data~\cite{AM98} within the
LO DGLAP approach provides a reasonable description of the high-energy
STAR data, if their overall normalization is scaled up by at least a
factor of two. This might indicate that the Sivers effect is
significantly stronger than thought so far and that small-$x$ effects
do not cancel in the ratio, by affecting the slope of the
cross section. More quantitative statements about the Sivers functions
must await data with smaller errors and over some range of energies.

\begin{acknowledgments}
We thank Werner Vogelsang for useful discussions regarding the DGLAP results,
and Mauro Anselmino, Umberto D'Alesio and Francesco Murgia for valuable
comments. 
The research of D.B.\ has been made possible by financial support 
from the Royal Netherlands Academy of Arts and Sciences and in
addition, is part of the research programme of the `Stichting voor
Fundamenteel Onderzoek der Materie (FOM)', which is financially supported
by the `Nederlandse Organisatie voor Wetenschappelijk Onderzoek (NWO)'.

\end{acknowledgments}

\end{document}